\documentclass[aps,pre,twocolumn]{revtex4-1}
\usepackage{amsfonts,amssymb,amsmath,bm}
\usepackage{graphicx,epsfig,color}

\begin{document}

\title{Irreversibility of the two-dimensional enstrophy cascade}
\author{E. Piretto$^1$, S. Musacchio$^2$ and G. Boffetta$^1$}

\affiliation{
$^1$Department of Physics and INFN, Universit\`a di Torino,
via P. Giuria 1, 10125 Torino, Italy \\
$^2$Universit\'e de Nice Sophia Antipolis, CNRS, LJAD, UMR 7351, 
06100 Nice, France
}

\date{\today}

\begin{abstract}
We study the time irreversibility of the direct cascade in two-dimensional
turbulence by looking at the time derivative of the square vorticity 
along Lagrangian trajectories, a quantity which we call 
{\it metenstrophy}.
By means of extensive numerical simulations we measure the time 
irreversibility from the asymmetry of the PDF of the metenstrophy
and we find that it increases with the Reynolds number of the
cascade, similarly to what found in three-dimensional turbulence. 
A detailed analysis of the different contributions to the 
enstrophy budget reveals a remarkable difference with respect to what
observed for the direct cascade, in particular the role of the 
statistics of the forcing to determine the degree of irreversibility.
\end{abstract}


\maketitle

\section{Introduction}
\label{sec1}

Although the time reversibility of the Navier-Stokes equations is broken by the
viscous forces, it is not restored in the limit of vanishing viscosity. 
Indeed in this limit one obtains the regime of fully developed turbulence,
characterized by an irreversible flux of
energy from the large scales to the small scales where it is dissipated. 
The irreversibility of the energy flux in three-dimensional turbulence 
is responsible for the asymmetry of the {\it two-point} statistical 
observables, either in a fixed Eulerian reference frame as in the 
case of velocity structure functions which 
display a negative third moment \cite{frisch1995turbulence}, or for
the Lagrangian evolution of pairs of trajectories 
\cite{falkovich2013single}.

Recently, it has been shown on the basis of laboratory experiments and 
numerical simulations, how irreversibility in turbulence manifests at
the level of {\it single-point} observable \cite{xu2014flight}. 
By looking at the evolution of the energy along a fluid trajectory, it
has been shown that the particle acquires (kinetic) energy on a long time 
scale and loses it on a short time scale. This reflects the fact that energy
is injected in the flow by an external forcing at large, slow scales
and is dissipated by viscosity at small, fast scales \cite{pumir2016single}.
As a consequence, although in stationary conditions 
the mean temporal increment of energy vanishes, the full statistics is
not time-reversible and odd moments of energy increments are different from 
zero \cite{xu2014flight}.
Time irreversibility can be quantified by looking at the statistics of 
the power along a trajectory $p(t)=d/dt (v^2(t)/2)$ and it has been
shown that the third moment 
$\langle p^3 \rangle$ is negative and increases with the control parameter,
the Reynolds number, of the flow. 
Similar results have been found for the inverse cascade of energy in 
two-dimensional turbulence, which is characterized by the same 
scaling law of 3D turbulence \cite{boffetta2012two} and in 
compressible turbulence \cite{grafke2015time}.
Despite the fact that in 2D the energy flows towards the 
large scales (instead of the small scales), time irreversibility manifests
in two-dimensional turbulence as in 3D, with a negative skewness of the 
power computed along a Lagrangian trajectory \cite{xu2014flight}.

In this paper, we study the time irreversibility of the {\it direct cascade}
in two-dimensional turbulence, characterized by an enstrophy (mean square
vorticity) flow towards small scales, by studying the statistics 
of what we call {\it metenstrophy}, i.e. the time derivative of the 
enstrophy along a trajectory. The main motivation for this 
work is that the direct cascade is characterized by a single characteristic
time \cite{boffetta2012two} and therefore spatial separation
(between injection and dissipation scales) does not corresponds in a 
simple way to 
time-scale separation. This make this turbulence completely different
from the energy cascades (direct and inverse) studied in \cite{xu2014flight}.
On the basis of direct numerical simulations at different Reynolds number,
we find that also in this case single-point statistics breaks time 
reversal symmetry and that the degree of irreversibility grows with the 
Reynolds numbers of the flow. 
A part this similarity, the picture which emerges from
the direct enstrophy cascade is very different from what observed in the 
energy cascade, as here the terms that contributes to the enstrophy
balance are all local and pressure gradient, which is responsible for the 
transfer of energy from slow to fast particles in 3D 
\cite{pumir2014redistribution} is absent.

The remaining of this paper is organized as follow. 
In Section~\ref{sec2} we provide a brief survey of two-dimensional turbulence, 
introducing the equations and the main quantities which will be discussed in the paper. 
In Section~\ref{sec3} we report the results of our numerical simulations. 
Section~\ref{sec4} is devoted to the discussion of our findings. 

\section{Theoretical background}
\label{sec2}
We consider the two-dimensional , 
incompressible Navier-Stokes equation for the vorticity field 
$\omega = \partial_x u_y - \partial_y u_x$
in a double periodic square box 
of dimension $L \times L$
\begin{equation}
\partial_t \omega + {\bm u} \cdot {\bm \nabla} \omega =
\nu \nabla^2 \omega - \alpha \omega + f
\label{eq:2.1}
\end{equation}
where $\nu$ is the kinematic viscosity, $\alpha$ the friction coefficient
and $f({\bm x},t)$ is an external forcing needed to sustain a stationary
state.
In the inviscid ($\nu=\alpha=0$) unforced ($f=0$) limit (\ref{eq:2.1})
has two conserved quantities: 
kinetic energy $E=(1/2) \langle u^2 \rangle$ 
and enstrophy $Z=(1/2) \langle \omega^2 \rangle$. 
Here $\langle ... \rangle$ represent average over the $L^2$ domain). 
The forcing term $f$ inject energy and enstrophy at a characteristic
scale in the system from which energy and enstrophy are transported 
towards large and small scales respectively, generating the double
cascade predicted by Kraichnan $50$ years ago 
\cite{kraichnan1967inertial,boffetta2012two}.
The inverse cascade of energy is characterized by Kolmogorov-like spectrum
with close-to-Gaussian statistics \cite{boffetta2010evidence}
and its time-reversal properties have been object of previous works
\cite{xu2014flight,pumir2014redistribution}. 
In the direct cascade to small scales, enstrophy is transferred at a rate
$\eta$ from the forcing scales $\ell_{f}$ down to the dissipative scales
$\ell_{\nu} \sim \nu^{1/2} /\eta^{1/6}$ generating a power spectrum
with exponent close to the dimensional prediction $-3$
\cite{maltrud1991energy,borue1993spectral,gotoh1998energy,
belmonte1999velocity,lindborg2000kinetic,
pasquero2002stationary,chen2003physical,rivera2003energy}.
The ratio of these scales defines the Reynolds number of 
the cascade as $Re=\eta^{1/3} \ell_{f}^{2}/\nu = (\ell_f/\ell_{\nu})^2$.

In presence of forcing and dissipation, the time derivative of the 
local square vorticity along a trajectory, which will be called 
{\it metenstrophy} $q$, is given by
\begin{equation}
q \equiv {D \over Dt} {1 \over 2} \omega^2 = q_{\nu} + q_{\alpha} + q_{f}
\label{eq:2.2}
\end{equation}
where with obvious notation we have introduced 
$q_{\nu}=\omega \nu \nabla^2 \omega$, $q_{\alpha}=-\alpha \omega^2$ and 
$q_{f}=\omega f$. In stationary conditions we have $\langle q \rangle=0$
and the enstrophy balance reads $\langle q_f \rangle = - \langle q_{\nu}
\rangle - \langle q_{\alpha} \rangle$ where the viscous dissipation is 
equal to the enstrophy flux $-\langle q_{\nu} \rangle=\eta$, while 
the large scale enstrophy dissipation $\langle q_{\alpha} \rangle$ is
negligible for large Reynolds numbers \cite{boffetta2010evidence}.

We observe that in the decomposition (\ref{eq:2.2}) all terms are local,
involving products of the vorticity field and its derivatives. This is
the main difference with respect to the balance for the energy cascade
in which a non-local term given by the pressure gradient is present. 
Although the pressure forces on average do not contribute 
to the kinetic energy balance, in 3D they are responsible for the 
redistribution of energy from slow to fast particles and for the asymmetry
of the PDF of energy power.
The absence of the analogous of the pressure forces in (\ref{eq:2.2})
suggests that the statistics of metenstrophy in the 2D direct cascade 
will be very different from the statistics of power in 3D.

\section{Numerical results}
\label{sec3}

We have integrated the Navier-Stokes Equations~(\ref{eq:2.1}) 
on a doubly periodic square domain of size $L=2\pi$ 
at resolution $N^2 = 1024^2$, 
by means of a standard fully-dealiased pseudospectral code, 
with 4th-order Runge-Kutta scheme with implicit integration 
of the linear dissipative terms. 
In order to avoid contamination of the enstrophy cascade
\cite{boffetta2002intermittency}
the linear friction term $- \alpha \omega$ 
has been replaced by an ipo-friction term $-\alpha \nabla^{-4} \omega$
which confines the dissipation to the largest scales.

Considering that the variations of the enstrophy along a Lagrangian trajectory 
are due only to the contributions of the forcing and dissipation, 
we have performed two sets of simulations aimed to 
investigate separately the effects of the viscous dissipation 
and of the external force on the metenstrophy statistics. 

In the following, all the results are non-dimensionalized with the enstrophy flux $\eta$ 
and with the characteristic time of the flow defined as 
$\tau_{\omega} = \eta^{-1/3}$. 

\subsection{Dependence on Reynolds}

We performed a first set of simulation (Set A) 
to investigate the dependence of the statistics on the Reynolds number 
by gradually reducing the viscosity 
and keeping fixed the forcing.  
In these simulations we use a deterministic, time-independent force 
$f(x,y) = F \sin(k_f x)\sin(k_f y)$. 
The forcing scale is defined as $\ell_f = 2 \pi /(k_f\sqrt{2})$ and
the parameters of the simulations are reported in Table~\ref{tab:1}. 
The statistics is computed over $2000$ independent vorticity fields, 
sampled every $1.2 \tau_\omega$. 

\begin{table*}[ht!]
\begin{tabular}{|ccccccc|}
\hline
\hline
$Re$ & $\nu$ & $-\langle q_\nu\rangle $ & $- \langle q_\alpha \rangle$ & $ \langle q_f \rangle$ & $E$ & $Z$ \\
\hline
$7.8 \cdot 10^2$ & $        10^{-3}$ & $2.48 \cdot 10^{-1}$ & $2.4 \cdot 10^{-2}$ & $2.72 \cdot 10^{-1}$ & $1.06 \cdot 10^{-1}$ & $1.99$ \\
$1.5 \cdot 10^3$ & $5 \cdot 10^{-4}$ & $2.24 \cdot 10^{-1}$ & $2.8 \cdot 10^{-2}$ & $2.53 \cdot 10^{-1}$ & $1.30 \cdot 10^{-1}$ & $2.53$ \\
$3.7 \cdot 10^3$ & $2 \cdot 10^{-4}$ & $2.13 \cdot 10^{-1}$ & $3.4 \cdot 10^{-2}$ & $2.47 \cdot 10^{-1}$ & $1.68 \cdot 10^{-1}$ & $3.62$ \\
$7.3 \cdot 10^3$ & $1 \cdot 10^{-4}$ & $2.07 \cdot 10^{-1}$ & $3.7 \cdot 10^{-2}$ &$ 2.44 \cdot 10^{-1}$ & $1.38 \cdot 10^{-1}$ & $4.25$ \\
$1.5 \cdot 10^4$ & $5 \cdot 10^{-5}$ & $2.05 \cdot 10^{-1}$ & $4.0 \cdot 10^{-2}$ & $2.46 \cdot 10^{-1}$ & $2.27 \cdot 10^{-1}$ & $5.75$ \\
\hline
\hline
\end{tabular}
\label{tab:1}
\caption{Parameters of the simulations with deterministic forcing (Set A). 
The amplitude of the force is $F=1$ and the forcing wavenumber
$k_f=4$. The ipo-friction coefficient is $\alpha=1$.}
\end{table*}

\begin{figure}[ht!]
\includegraphics[width=\columnwidth]{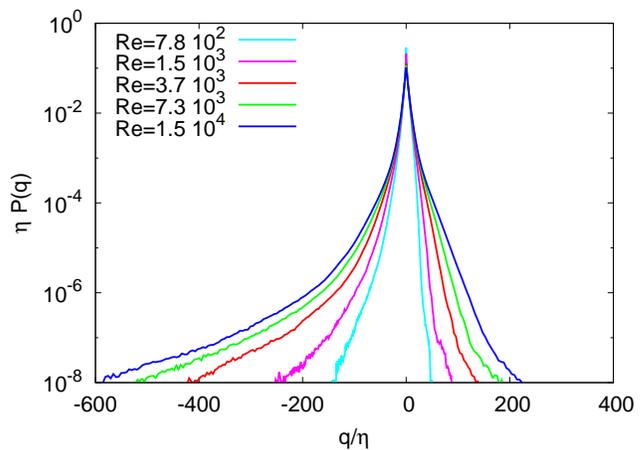}
\caption{(Color online). 
PDFs of metenstrophy $q$ at increasing $Re$ (from the inner to the outer PDF)}
\label{fig:1}
\end{figure}

In Figure~\ref{fig:1} we show the probability distribution functions (PDFs) 
of the metenstrophy $q$ for the simulations of the Set A at different Reynolds numbers.
Even if the mean value of $q$ vanishes, because of the statistical 
time-stationarity, 
the full statistics reveals a noticeable violation of the time symmetry. 
In particular, at increasing $Re$ we observe the development 
of a large left tail in the PDFs and the third moment of $q$ become negative.

\begin{figure}[ht!]
\includegraphics[width=\columnwidth]{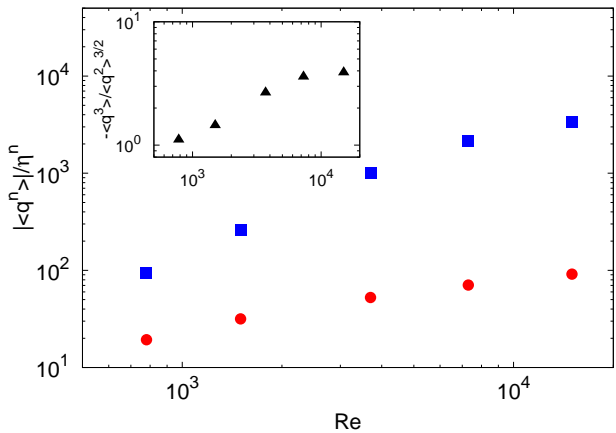}
\caption{(Color online). 
Second (red circles) and third (blue square) moments of the 
PDFs of metenstrophy $q$ as a function of $Re$.
Inset: Skewness of metenstrophy as a function of $Re$.}
\label{fig:2}
\end{figure}

Both the second moment $\langle q^2 \rangle$ and the third moment 
$\langle q^3 \rangle$ of the distribution grow monotonically with $Re$ 
(see Figure~\ref{fig:2}), similarly to what observed for the statistics
of power in 3D turbulence \cite{xu2014flight}. At variance with the 3D case, 
here we are unable to find a clear power-law scaling for the two moments
but this could still be due to finite Reynolds effect.

The skewness $S = \langle q^3 \rangle / \langle q^2 \rangle^{3/2}$ 
provides a suitable measure of the irreversibility. 
The results of our numerics show that it increases with $Re$ 
(see Figure~\ref{fig:2}) and suggest a possible saturation to a constant 
value for large values of $Re$. We remark that the saturation of the skewness
to a constant value has been observed for the power in 3D turbulence.

\begin{figure}[ht!]
\includegraphics[width=\columnwidth]{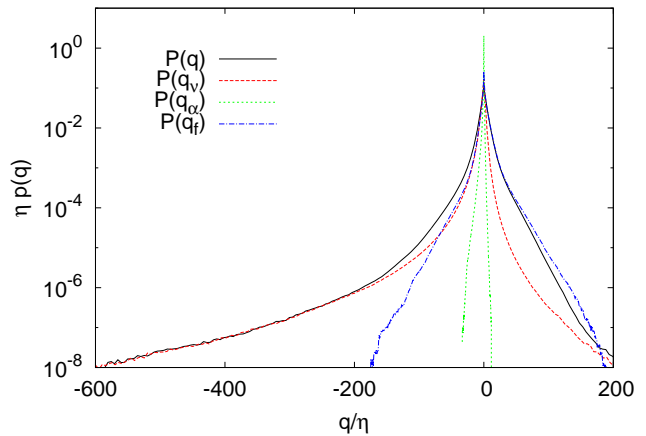}
\caption{
(Color online).
PDF of the metenstrophy $q$ for the case $Re=1.5 \cdot 10^4$ (back solid line), 
and of the different contributions: 
viscosity $q_\nu$ (red, dashes line), 
ipofriction $q_\alpha$ (green, dotted line), 
forcing $q_f$ (blue, dash-dotted line). 
}
\label{fig:3}
\end{figure}

In order to understand which physical process 
is responsible for the breaking of the time-symmetry, 
we have analyzed the different contributions to the metenstrophy, 
due to the forcing $q_f$, the viscous dissipation $q_\nu$ and the ipo-friction $q_\alpha$. 
In Figure~\ref{fig:3} we compare the PDF of $q$ for the run at $Re=1.5 \cdot 10^4$
with the PDFs of $q_f$, $q_\nu$ and $q_\alpha$.
We find that the large left tail of the PDF of $q$,  
is dominated by local events of intense viscous dissipation, 
which can be $600$ times more intense than their mean.
Conversely, the statistics of the forcing contributions $q_f$ 
is more symmetric, 
and it prevails in the right tail of $P(q)$. 
As expected, the contributions of the ipo-friction are negligible on 
the statistics of $q$. 

\begin{figure}[ht!]
\includegraphics[width=\columnwidth]{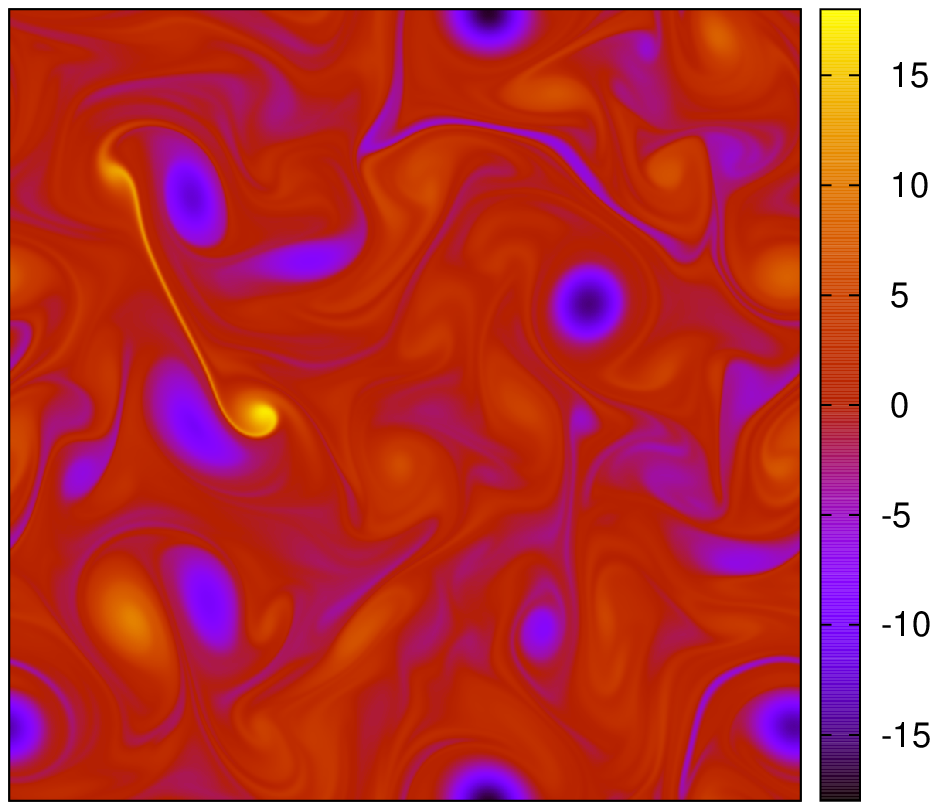}
\includegraphics[width=\columnwidth]{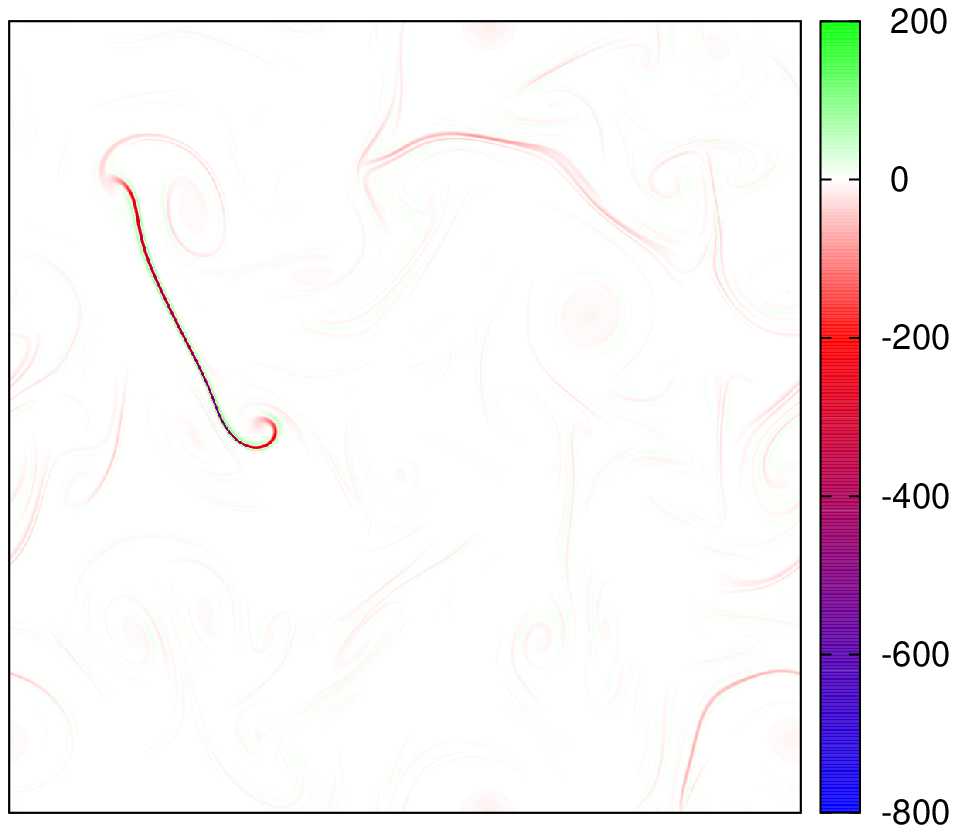}
\caption{
(Color online). Snapshots of the vorticity field $\omega$ (upper plot) 
and viscous enstrophy dissipation rate $q_{\nu}$ (lower plot) at the
same time for the simulation at $Re=3.7 \cdot 10^3$. 
}
\label{fig:4}
\end{figure}

We have observed a signature of the presence of strong dissipative events 
also in the statistical convergence of the moments of $q$, 
which displays abrupt changes of the averages during the evolution 
of the system. 
A visual inspection of the vorticity field at the time these events occurs, 
reveals the presence of extremely intense and tiny filaments of vorticity 
(see Figure~\ref{fig:4}). 
These vorticity filaments, which are generated by the chaotic stretching 
of the direct enstrophy cascade, 
causes localized events of strong viscous dissipation, 
which are clearly visible in the corresponding field $q_\nu$
\cite{bracco2010reynolds}.

In summary, the results of this set of simulations reveal a significant
breaking of the time-asymmetry in the statistics of the metenstrophy.  
The irreversibility increases with Reynolds, 
and it is intrinsically related to the 
chaotic-stretching nature of the direct enstrophy cascade, 
which produces tiny and intense 
filaments of vorticity localized both in space and time. 

\subsection{Dependence on the forcing correlation time}

Considering that the intense dissipative events 
must be balanced on average by the forcing, 
it is natural to suppose that the statistics of the metenstrophy 
cannot be universal with respect to the forcing itself.  
We addressed this issue with a second set of simulations (Set B), 
in which we keep fixed the viscous dissipation 
and we have changed the time-correlation $\tau_f$ of the external forcing. 
For this purpose, 
we forced all the wavenumbers in the shell $k \in [k_{f1}, k_{f2}]$
with independent stochastic Ornestein-Uhlenbeck processes 
$d f_{t,{\bm k}} = -(1/\tau_f) f_{t,{\bm k}} dt +\sqrt{2F} d W_{t,{\bm k}}$, 
where $W_{t,{\bm k}}$ are independent Wiener processes.  
The amplitude $F$ of the forcing has been tuned to obtain (a posteriori) 
similar enstrophy fluxes in the simulations with different $\tau_f$. 
We define the mean square forcing wavenumber as 
$k_F^2 = \langle |(\bm{k}|^2 \rangle_{k \in [k_{f1}, k_{f2}]}$
and the forcing scale as $\ell_f = 2 \pi /k_f$. 
The parameters of this second set of simulations are reported in
Table~\ref{tab:2}.
Also in this case, the statistics is computed over $2000$ independent 
vorticity fields, 
sampled every $1.2 \tau_\omega$. 

\begin{table*}
\begin{tabular}{|ccccccccc|}
\hline
\hline
$Re$ & $\tau_f \eta^{1/3}$ & $\tau_f$ & $F$ & $-\langle q_\nu\rangle $ & $-\langle q_\alpha \rangle$ & $ \langle q_f \rangle$ & $E$ & $Z$ \\
\hline
$1.60 \cdot 10^4$ & $0.15$ & $0.25$ & $2.5 \cdot 10^{-1}$ & $2.08 \cdot 10^{-1}$ & $3.8 \cdot 10^{-2}$ & $2.50 \cdot 10^{-1}$ & $1.60 \cdot 10^{-1}$ & $3.52$ \\
$1.55 \cdot 10^4$ & $0.29$ & $$0.50 & $5.8 \cdot 10^{-2} $ & $1.89 \cdot 10^{-1}$ & $3.5 \cdot 10^{-2}$ & $2.25 \cdot 10^{-1}$ & $1.55 \cdot 10^{-1}$ & $3.45$\\
$1.63 \cdot 10^4$ & $0.60$ & $1.00$ & $2.0 \cdot 10^{-2}$ & $2.18 \cdot 10^{-1}$ & $4.1 \cdot 10^{-2}$ & $2.60 \cdot 10^{-1}$ & $1.80 \cdot 10^{-1}$ & $4.05$\\
$1.55 \cdot 10^4$ & $1.15$ & $2.00$ & $5.7 \cdot 10^{-3}$ & $1.87 \cdot 10^{-1}$ & $3.7 \cdot 10^{-2}$ & $2.25 \cdot 10^{-1}$ & $1.68 \cdot 10^{-1}$ & $3.90$\\
$1.57 \cdot 10^4$ & $2.91$ & $5.00$ & $1.8 \cdot 10^{-3}$ & $1.96 \cdot 10^{-1}$ & $4.0 \cdot 10^{-2}$ & $2.37 \cdot 10^{-1}$ & $1.92 \cdot 10^{-1}$ & $3.53$\\
\hline
\hline
\end{tabular}
\label{tab:2}
\caption{Parameters of the simulations with Ornestein-Uhlenbeck forcing (Set B).  
The wavenumber forcing shell is $5 \leq k \leq 6$.  
The viscosity is $\nu=5 \cdot 10^{-5}$ 
and the ipo-friction coefficient is $\alpha=1$. 
}
\end{table*}

\begin{figure}[h!]
\includegraphics[width=\columnwidth]{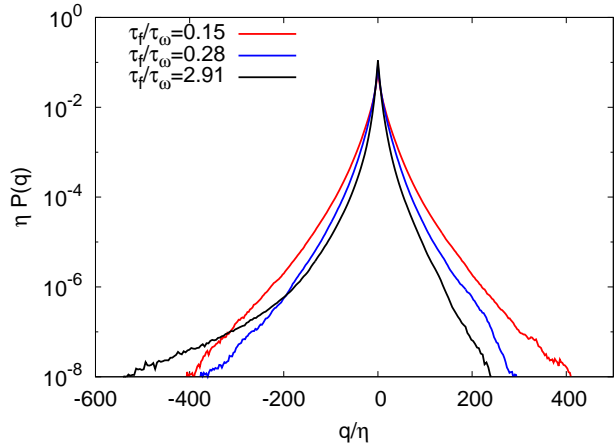}
\caption{(Color online). 
PDFs of metenstrophy $q$ for different correlation times $\tau_f$}
\label{fig:5}
\end{figure}

The PDFs of the metenstrophy computed in simulations with different forcing  
are shown in Figure~\ref{fig:5}. 
In the simulations with a forcing with the long correlation time $\tau_f$ 
we observe PDFs characterized by a strong asymmetry, due to a pronounced 
left tail, similar to what observe for the stationary forcing of the Set A.
However, we find that the asymmetry reduces 
as we reduce the correlation time of the forcing. 

\begin{figure}[ht!]
\includegraphics[width=\columnwidth]{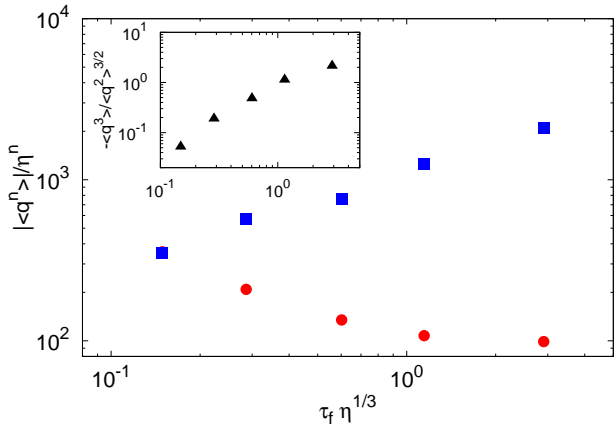}
\caption{(Color online). Second (red circles) and third (blue square) 
moments of the PDFs of metenstrophy $q$ as a function of $\tau_f$.
Inset: Skewness of metenstrophy as a function of $\tau_f$.}
\label{fig:6}
\end{figure}

The simmetrizzation of the PDF is accompained by a broadening of its tails. 
This is well captured by the dependence of the second and third moment 
on $\tau_f$ shown in Figure~\ref{fig:6}. 
As the correlation time is reduced, we observe an increase of the second moment, 
which corresponds to the broadening of the PDF's tails, 
and a reduction of the third moment of the distributions, 
which indicates the reduction of the asymmetry.  

The combined growth of $\langle q^2 \rangle$ 
and the decrease of $\langle q^3 \rangle$ 
results in a reduction of the skewness 
$S = \langle q^3 \rangle / \langle q^2 \rangle^{3/2}$ 
at reducing the correlation time of the forcing (see Figure~\ref{fig:6}). 

\begin{figure}[ht!]
\includegraphics[width=\columnwidth]{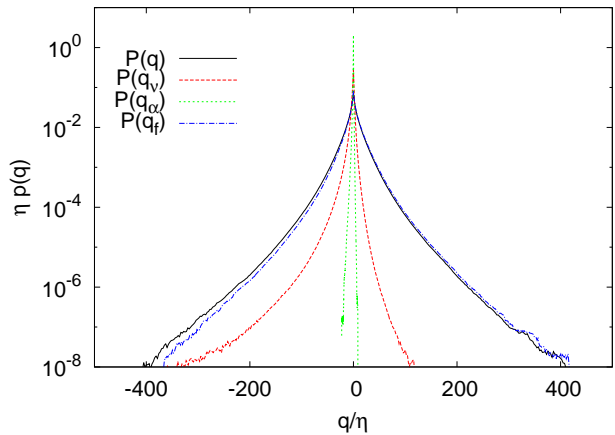}
\caption{
(Color online).
PDF of the metenstrophy $q$ for the case $\tau_f \eta^{1/3} = 0.15$ (back solid line), 
and of the different contributions: 
viscosity $q_\nu$ (red, dashes line), 
ipofriction $q_\alpha$ (green, dotted line), 
forcing $q_f$ (blue, dash-dotted line). 
}
\label{fig:7}
\end{figure}

The analysis of the different contribution to the metenstrophy, 
(the forcing $q_f$, the viscous dissipation $q_\nu$ and the ipo-friction 
$q_\alpha$)
reveals that the fluctuations of the forcing are the main responsible 
for the broadening of the tails observed 
in the case of short-correlated forcing (see Figure~\ref{fig:7}).
The PDF of the viscous dissipation $q_\nu$ displays a clear asymmetry, 
but its contribution to $P(q)$ is much weaker than that of the forcing. 

Our findings can be explained easily. 
Reducing the time-correlations of the forcing  
causes also a reduction of the correlation between 
the force field and the vorticity field. 
The mean, positive enstrophy input 
is therefore the result of cancellations between 
larger and larger positive and negative  
fluctuations of the input, which have no reason to be asymmetric.  
The broad, symmetric tails of $P(q_f)$ which develops in the limit 
$\tau_f \to 0$ 
overwhelm the asymmetric contributions of $P(q_\nu)$, 
originated by the generation of the tiny vorticity filaments. 

\section{Conclusions}
\label{sec4}

In this work we have investigated the statistics of the metenstrophy $q$, 
that is the time derivative of the enstrophy along
a Lagrangian trajectory, in a two-dimensional turbulent flows
in the regime of the direct enstrophy cascade
sustained by deterministic and stochastic forcing. 

The main result of our work is that the statistics of $q$ 
is characterized by a strong violation of the time-symmetry. 
The irreversibility increases with the Reynolds number, 
and it is deeply related to the mechanism of the direct enstrophy cascade, 
which generates tiny filaments of vorticity by means of chaotic stretching. 
At the viscous scales, these filaments causes intense events of enstrophy dissipation, 
therefore giving strong contributions to the left tail of the PDF of $q$.  

Being the results of balance between forcing and dissipation, 
the statistics of the metenstrophy is also dependent on the forcing mechanisms. 
In particular we have shown that the irreversibility is reduced 
in the case of stochastic forcing with short correlation time, 
whose broad and symmetric fluctuations overwhelms 
the asymmetric contributions of the viscous dissipation. 

The mechanism which causes the symmetry breaking 
is essentially the chaotic stretching of the flow. 
This suggests that our results can be extended also to other systems, 
in particular to the statistics of a scalar field 
transported a turbulent of a chaotic flow. 

More in general, the study of single point irreversibility in different
turbulent models will allow to build a general picture of possible
universal features of how time symmetry breaking in far from equilibrium
systems.


\bibliography{biblio}

\end{document}